\documentclass[twocolumn,secnumarabic,superscriptaddress,showpacs,fleqn,showkeys, amssymb, nobibnotes, aps, prb]{revtex4-2}
\usepackage{graphicx}
\usepackage{dcolumn}
\usepackage{bm}
\usepackage{xcolor}
\usepackage{afterpage}

\begin{document}

\title{Coexistence of local structural heterogeneities and long-range ferroelectricity in Pb-free (1-x)Ba(Ti$_{0.8}$Zr$_{0.2}$)O$_3$-x(Ba$_{0.7}$Ca$_{0.3}$)TiO$_3$ ceramics}
\author{K. Dey}
 \address{UGC DAE Consortium for Scientific Research, University Campus, Khandwa Road, Indore 452001, India.}
 \author{A. Ahad}
 \address{Department of Physics, Aligarh Muslim University, Aligarh 202002, India.}
 \author{K. Gautam}
 \address{UGC DAE Consortium for Scientific Research, University Campus, Khandwa Road, Indore 452001, India.}
\author{A. Tripathy}
 \address{UGC DAE Consortium for Scientific Research, University Campus, Khandwa Road, Indore 452001, India.}
\author{S. S. Majid}
 \address{Department of Physics, Aligarh Muslim University, Aligarh 202002, India.}
 \thanks{Now at Institute of Earth Sciences, Academia Sinica, Taipei 115201, Taiwan.}
 \author{S. Francoual}
 \address{Deutsches Elektronen-Synchrotron (DESY), Notkestrasse 85, D-22607 Hamburg, Germany.}
 \author{C. Richter}
 \address{The European Synchrotron Radiation Facility, 71 Avenue des Martyrs, 38000 Grenoble, France.}
 \thanks{Now at The Leibniz-Institut für Kristallzüchtung, Max-Born-Straße 2, 12489 Berlin, Germany.}
 \author{M. N. Singh}
 \address{Synchrotrons Utilization Section, Raja Ramana Centre for Advance Technology, Indore 452013, India.}
 \author{A. Sagdeo}
 \address{Synchrotrons Utilization Section, Raja Ramana Centre for Advance Technology, Indore 452013, India.}
 \address{ Homi Bhabha National Institute, Training School Complex, Anushakti Nagar, Mumbai, 400094, India.}
 \author{E. Welter}
 \address{Deutsches Elektronen-Synchrotron (DESY), Notkestrasse 85, D-22607 Hamburg, Germany.}
 \author{N. Vittayakorn}
 \address{Advanced Materials Research Unit, Faculty of Science, King Mongkut's Institute of Technology Ladkrabang, Bangkok, Thailand 10520.}
\author{V. G. Sathe}
 \address{UGC DAE Consortium for Scientific Research, University Campus, Khandwa Road, Indore 452001, India.}
 \author{R. Rawat}
 \address{UGC DAE Consortium for Scientific Research, University Campus, Khandwa Road, Indore 452001, India.}
\author{D. K. Shukla}%
 \thanks{Corresponding Author: dkshukla@csr.res.in}
\address{UGC DAE Consortium for Scientific Research, University Campus, Khandwa Road, Indore 452001, India.}
\begin{abstract}
Environmentally benign (1-x)Ba(Ti$_{0.8}$Zr$_{0.2}$)O$_3$-x(Ba$_{0.7}$Ca$_{0.3}$)TiO$_3$ (BZT-BCT) ceramics are promising materials due to their remarkable high piezoresponse [Liu and Ren, Phys. Rev. Lett. \textbf{103}, 257602 (2009)]. In this Letter, by focusing on local and average structure in combination with macroscopic electromechanical and dielectric measurements we demonstrate the structure property relationship in the tetragonal BZT-BCT ceramic. During high-temperature cubic to tetragonal phase transformation, polar nanoregions are manifested through the spontaneous volume ferroelectrostriction at temperatures below $\sim$ 477 K. Temperature-dependent local structural investigations across the Zr K edge extended x-ray absorption fine structure spectroscopy reveal an anomalous collaboration between the ZrO$_{6}$ and TiO$_6$ octahedra. These octahedra compromise their individuality during polarization development. The presence of domains of submicron size embedded inside the macroscopic ferroelectric regions below T$_{m}$, as well as their hierarchical arrangement, is observed by piezo-response force microscopy. Effects of the existence of the structural/polar heterogeneities below T$_{m}$ are observed also when polarizibilities of the poled and the unpoled samples are compared; the poled sample is found to be more susceptible to the electric field. In addition, by using electric field dependent x-ray diffraction studies we also show that this ceramic under field exhibits reduction of tetragonal distortion, which is consistent with earlier reports.          
\end{abstract}       
\maketitle
\section{Introduction}
  Investigations on lead-free piezoelectric materials in the last two decades have largely intensified due to environmental concerns. Liu \& Ren \cite{liu2009large} demonstrated PZT (PbZr$_{x}$Ti$_{1-x}$O$_3$) like `morphotropic phase boundary' (MPB) starting from a triple point in BZT-BCT (1-x)Ba(Ti$_{0.8}$Zr$_{0.2}$)O$_3$-x(Ba$_{0.7}$Ca$_{0.3}$)TiO$_3$ which possesses ultrahigh piezoresponse (d$_{33}$) values as high as up to 560-620 pC/N) among all lead free compositions. From Liu \& Ren \cite{liu2009large} and Nahas \cite{nahas2017microscopic}\textit{et al.,} high piezoresponse has been attributed to isotropic energy flattening along with miniaturization of domains towards the triple point \cite{gao2016phase}. In addition, domain wall motion was reported to be the dominant extrinsic contribution to high piezoresponse \cite{gao2014major}. Electric field dependent x-ray diffraction (XRD) measurements \cite{tutuncu2014domain} revealed that spontaneous lattice strain reduction enhances domain wall motion in BZT-BCT ceramics. In morphotropic composition (BZT-50BCT) Brajesh \textit{et al.,} \cite{brajesh2015relaxor, brajesh2016structural} showed phase coexistence at room temperature (Tetragonal + Rhombohedral + Orthorhombic) and concluded that electric field induced metastable phases contribute significantly in high piezoresponse. They also showed relaxor ferroelectric behavior and an ergodic relaxor to normal ferroelectric transformation at $\sim$ T$_{m}$. Relaxor ferroelectricity i.e., the existence of polar nanoregions (PNRs), a nanoscale inhomogeneity, has been prototypical of highest piezoresponse in Pb-based materials \cite{cross1987relaxor, manley2016giant, park1997ultrahigh}. However, in BZT-BCT ceramics, among all posibilities of high piezoresponse, discussed so far, thorough validation of PNRs is yet missing. 
 
In Pb-based perovskite relaxor ferroelectrics, the quenched compositional disorder induced by hetrovalent substitution at the B site (e.g. Pb(Mg$_{1/3}$Nb$_{2/3}$)O$_{3}$-PbTiO$_{3}$, Pb(Zn$_{1/3}$Nb$_{2/3}$)O$_{3}$-PbTiO$_{3}$) generates random fields which break long range ferroelectricity and frustrated interactions among dipoles start to form nanoscale correlations (PNRs) at a higher temperature which is known as the characteristic Burn's temperature, T$_B$ \cite{bokov2006recent,burns1983glassy}. Furthermore, the effect of the Pb positional disorder and its coupling with different BO$_6$ octahedra also plays a critical role \cite{perrin2000neutron,halasyamani2004asymmetric}. The PNRs exhibit a large distribution of relaxation times. Divergence of the longest relaxation time takes place at a characteristic temperature, T$_{f}$, the freezing temperature \cite{pirc2007vogel}. In the ergodic relaxor phase, between two characteristic temperatures (T$_B$ \& T$_{f}$) PNRs have intrinsic dynamic character and after some percolation limit they trasfer into ferroelectric domains just below T$_{f}$ where macroscopic symmetry changes. However, in the case of canonical relaxors (for e.g. PMN), the macroscopic symmetry remains the same during ergodic to nonergodic transition, and PNRs show glassy interactions  among dipoles in the non-ergodic relaxor phase which transform into ferroelectric domains upon the application of sufficiently large electric field \cite{bokov2012dielectric}.

BaTiO$_3$-based compositions (e.g. BaZr$_{x}$Ti$_{1-x}$O$_3$, BaSn$_{x}$Ti$_{1-x}$O$_3$, BaHf$_{x}$Ti$_{1-x}$O$_3$) exhibit broad dielectric permittivity maxima with frequency dispersion and characteristic temperatures (T$_B$ \& T$_{f}$) similar to those Pb-based relaxors \cite{simon2000relations,huang2020multifunctional}, and in compositions also the possible existence of PNRs have been reported \cite{liu2007structurally,xie2012static}. However, the origin of PNRs in homovalent substituted BaTiO$_3$-based compositions is different than in Pb-based (as explained above) and challenging to unravel. A pair distribution function study \cite{pramanick2018stabilization} of BZT (BaZr$_{x}$Ti$_{1-x}$O$_3$) revealed dynamic insights and highlighted the role of Zr in slowing down the B site (Ti) dynamics (responsible for PNRs formation). Therefore, in order to have deeper insights into PNRs in BZT-BCT ceramics it is pivotal to have local structural information about the Zr/Ti ions.

We have chosen a tetragonal composition [(Ba$_{0.82}$Ca$_{0.18}$)(Ti$_{0.92}$Zr$_{0.08}$)O$_3$ (BZT-60BCT)] among the BZT-BCT systems mainly because it is far from the triple point \& morphotropic composition. In compositions near triple point and MPB structural instabilities due to the presence of metastable orthorhombic phases \cite{akbarzadeh2018quantum,brajesh2016structural} and nearly vanishing anisotropy induced domain wall motion \cite{tutuncu2014domain} may inhibit the study of intrinsic features of PNRs, which are hiding in these ceramics. 

In this letter, we demonstrate that, in BZT-60BCT below T$_{m}$, local structural heterogeneities coexist with long range ferroelectricity and this coexistence leads to hierarchical arrangements of micro and nano domains inside large ferroelectric domains. We also show direct evidences of PNRs above T$_{m}$. Our detail local structural studies of TiO$_{6}$ and ZrO$_{6}$ reveal a peculiar behavior of these two octahedra during polarization development and static disorder in Zr-O bonds below T$_m$.     

\section{EXPERIMENTS}
The ceramic BZT-60BCT has been synthesized by conventional solid state reaction using high-purity chemicals; BaCO$_3$ (99.9\%), CaCO$_3$ (99.9\%), TiO$_2$ (99.9\%), and ZrO$_2$ (99.7\%). Stoichiometrically mixed powders were thoroughly ball milled and calcined at 1300$^{\circ}$C for 4 h.  Sintering was performed at $\sim$1540$^{\circ}$C for 2 h in air. XRD measurement was done using Bruker D2 Phaser X-ray diffractometer utilizing Cu K$_{\alpha}$ radiation. Dielectric measurements were performed utilizing HIOKI LCR meter (IM 3536) and a CRYOCON temperature controller. Polarization-electric field (P-E) hysteresis loops were measured by Radiant precision premier II loop tracer. Temperature dependent XRD measurements were carried out at BL12 beamline at Indus II, RRCAT, India, by utilizing photons of $\sim$ 14.87 keV (0.834 $\AA$) in $\theta$-2$\theta$ geometry during the heating cycle from room temperature up to 575 K. The Raman measurements were carried out using a Horiba JY model HR-800 single spectrometer with high resolution grating and a CCD detector by using a 473 nm laser beam in backscattering geometry. Electric field dependent x-ray diffraction was performed using high-energy photons ($\sim$37 keV) in transmission geometry at beamline P09 at PETRA III at DESY (Hamburg, Germany). A PerkinElmer two-dimensional detector was used at a fixed distance of $\sim$ 650 mm. For poled measurements the sample was poled at $\sim$40$^{\circ}$C for 30 min at 1 kV/mm and then field cooled down to room temperature. After $\sim$ 24 hours of aging of the poled samples, piezoelectric charge coefficient d$_{33}$ was monitored by a commercial d$_{33}$ meter from APC. A similar poling strategy was adopted for all other poled measurements. EXAFS (extended X-ray absorption fine structure) measurements at the Zr \textit{K} edge in transmission and Ti \textit{K} edge XANES (X-ray absorption near edge structure) in fluorescence were collected at P65 beamline of PETRA-III synchrotron. Peizo-response force microscopy (PRFM) measurements were carried out on mirror finished polished surfaces. The sample was annealed at 350$^{\circ}$C for 2 hrs to remove stresses induced by polishing. PRFM measurements were carried out using a fiber interferometry based low temperature scanning probe microscope system from Attocube$^{TM}$, Germany. A PtSi-FM$^{TM}$ conducting cantilever from Nanosensors$^{TM}$ was used for scanning sample surface. Topography was measured in contact mode along with simultaneous measurement of piezo-response at an applied ac voltage of 10 V (peak to peak) at 41.33 kHz. 
\begin{figure}
\includegraphics[scale=0.34]{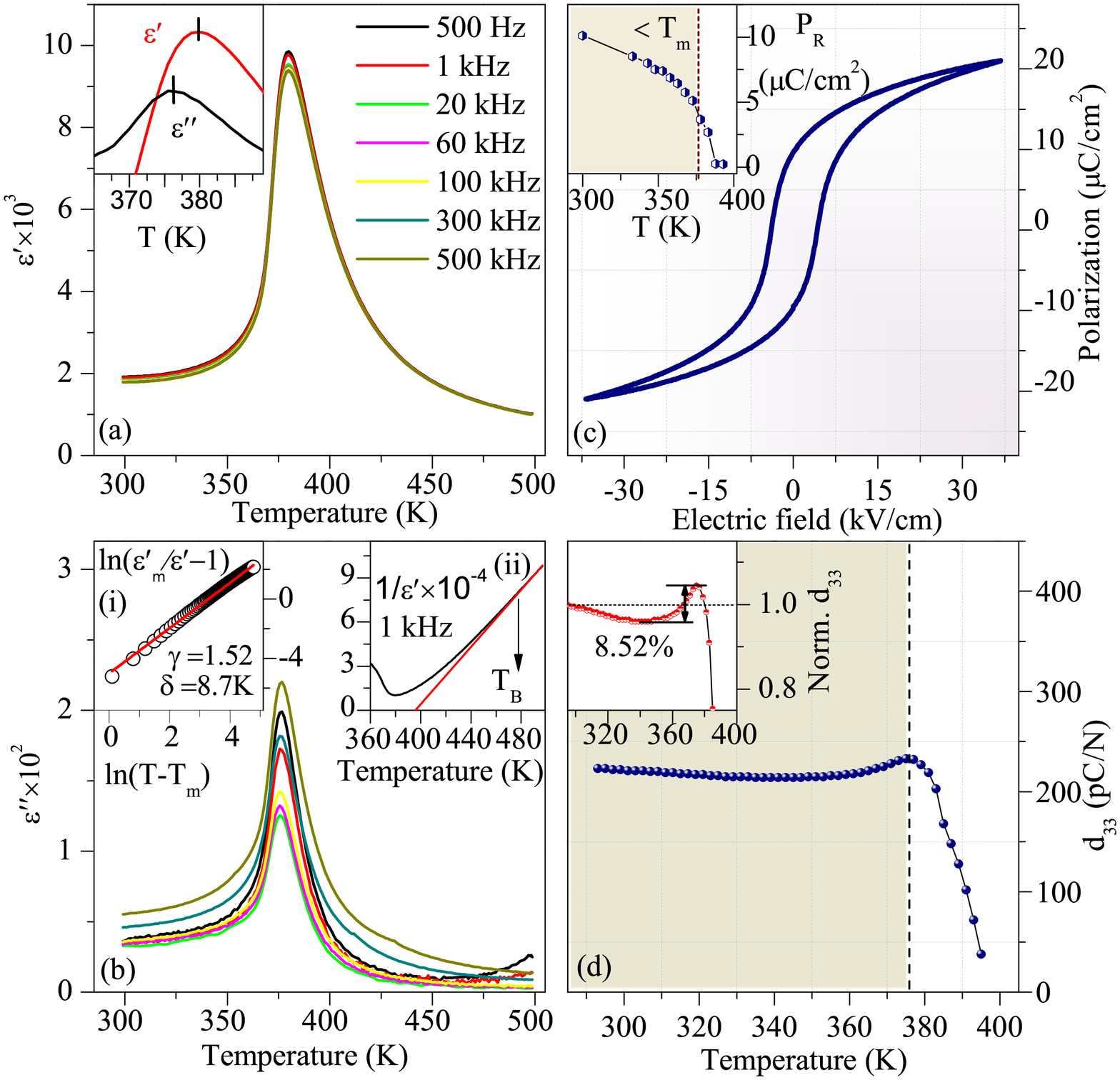}
\caption{\label{fig:epsart} Temperature dependence of (a) the real \& (b) imaginary parts of relative permittivity at different frequencies. The inset in (a) shows the zoomed portion of real and imaginary permittivities near the peak at 1 kHz. The bars indicate that T$_m$ for the imaginary part is at a lower temperature than that of the real part. Inset (ii) in (b) shows that the Curie-Weiss fit of the real permittivity indicates a deviation from the temperature marked T$_{B}$ ($\sim$ 477 K), and the inset (i) in (b) shows the linear fitting by using the empirical formula (see text). (c) Polarization- electric field hysteresis loop at room temperature; and inset shows the temperature dependence of the remnant polarization. (d) Temperature-dependent d$_{33}$; the inset shows normalized d$_{33}$ with temperature and arrow indicates the maximum percentage change in d$_{33}$.}
\label{Electrical_F_new}
\end{figure}
\begin{figure}
\includegraphics[scale=0.45]{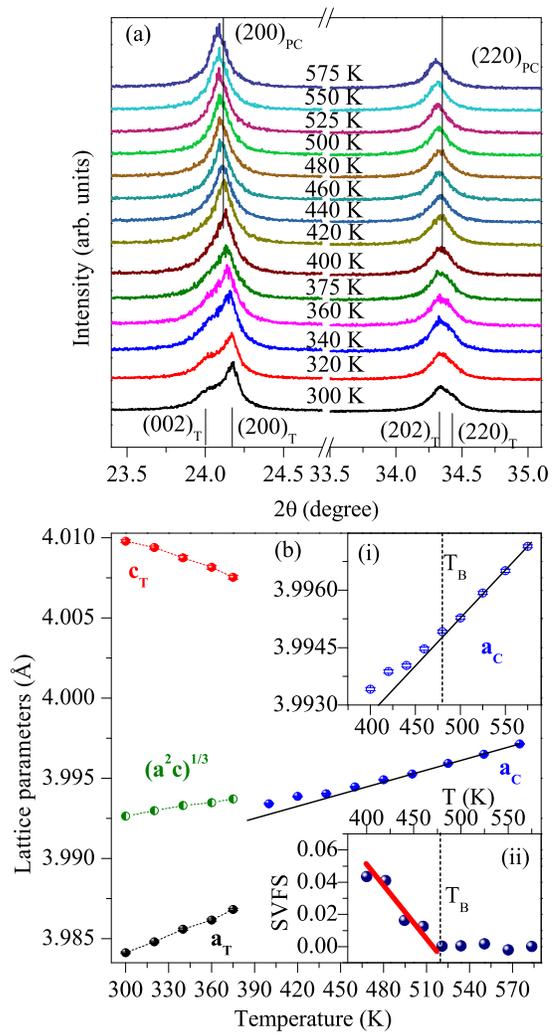}
\caption{\label{fig:epsart} (a) Temperature dependence of reflections (200)$_{PC}$ \& (220)$_{PC}$. (b) Temperature variation of lattice parameters in tetragonal and cubic structures. Insets (i) in (b) shows the zoomed portion of the cubic lattice parameters' temperature dependence. Inset (ii) in (b) shows the calculated spontaneous volume ferroelectrostriction (SVFS) with temperature.} 
\label{HTXRD1}
\end{figure}
\section{Results}
The room temperature tetragonal structure of BZT-60BCT has been confirmed using XRD (see Fig. S1 in the supplemental material \cite{supp1}). The real and imaginary parts of the dielectric permittivity show diffused dielectric maxima at $\sim$ 380 K (see Fig.~\ref{Electrical_F_new} (a) $\&$ (b)), signatures of relaxor ferroelectricity. Inset (ii) in Fig.~\ref{Electrical_F_new}(b) shows that the inverse permittivity deviates from the Curie-Weiss law at $\sim$ 477 K, which indicates a characteristic Burn's temperature, T$_{B}$ \cite{burns1983crystalline}. The diffusion exponent ($\gamma$) and diffuseness ($\delta$) of permittivity peak are calculated from the fit (see inset (i) in \ref{Electrical_F_new} (b)) using an empirical formula from reference \cite{bokov2003empirical},$\frac{\varepsilon_{m}}{\varepsilon}= 1+\frac{(T-T_{m})^{\gamma}}{2\delta^{2}}$, where T$_{m}$ is the peaking temperature at selected frequency. Obtained fitting parameters are $\gamma$ = 1.52 \& $\delta$ = 8.7 K. The polarization-electric field (P-E) hysteresis loop in Fig.~\ref{Electrical_F_new}(c) exhibits coercive field (E$_C$) $\sim$ 4 kV/cm and remanence $\sim$ 10 $\mu$C/cm$^{2}$. The temperature dependent remanence (see inset in Fig.~\ref{Electrical_F_new}(c)) is plotted from P-E loop measurements done at various temperatures \cite{supp1} and shows a non zero value of polarization even above T$_m$ $\sim$ 380 K. The observations of Burn's temperature and non-zero value of polarization above T$_{m}$ indicate the existence of PNRs at temperatures above T$_m$. A longitudinal piezoelectric coefficient (d$_{33}$) of $\sim$ 223 pC/N and its temperature insensitivity up to $\sim$ 375 K can be radially seen in Fig.~\ref{Electrical_F_new}(d); the inset shows normalized d$_{33}$ with temperature and its maximum variation is found to be only $\sim$ 8.52 \%, which is better than requirement ($\sim$ 10 $\%$) for applications.

To further substantiate the presence of PNRs and to have insights of these characteristic temperatures (T$_{B}$  \& T$_{m}$) observed in macroscopic electrical measurements we now focus on structural studies results obtained using synchrotron XRD measurements as a function of temperature (300 K to 575 K). Fig.~\ref{HTXRD1} (a) shows that high temperature pseudocubic (PC) reflections (200)$_{PC}$ \& (220)$_{PC}$ split with decreasing temperature in (200)$_{T}$/(002)$_{T}$ and (220)$_{T}$/(202)$_{T}$, respectively, signifying a structural phase transition from cubic to tetragonal.  

Rietveld refinement is done by using the Fullprof program \cite{rodriguez2001introduction} with space group P4mm from RT to 375 K and with the space group Pm-3m from 400 K up to 575 K (see Fig. S1 in supplemental material \cite{supp1}). Fig.~\ref{HTXRD1} (b) shows the temperature dependence of the refined lattice parameters. Transitions in lattice parameter trend clearly matches with the dielectric maxima temperature (T$_m$) (see Fig.~\ref{Electrical_F_new} (a)). Interestingly, the inset (i) in Fig.~\ref{HTXRD1}(b) shows deviation of the cubic lattice parameters at $\sim$ 477 K temperature (identified also as Burn's temperature T$_B$, see inset (ii) in Fig.~\ref{Electrical_F_new} (b)). These observed anomalies in structural parameters can be understood as follows. Ferroelectric order overcomes the decrease of anharmonic lattice phonon vibration and results into dilatation of cubic lattice parameter and this effect is termed as spontaneous volume ferroelectrostriction (SVFS) \cite{chen2015negative}. Here, in the cubic phase temperature behavior of the observed volume V$_{exp}$ is found to deviate at $\sim$ 477 K from its usual behavior V$_{nm}$. The SVFS is calculated by using the equation $\omega$$_s$= (V$_{exp}$-V$_{nm}$)/V$_{nm}$$\times$100$\%$, from references \cite{chen2013effectively,chen2015negative} shown in the inset (ii) of Fig.~\ref{HTXRD1} (b). Evolution of SVFS just below T$_B$ ($\sim$ 477 K) (inset (ii) in Fig.~\ref{HTXRD1} (b)) signifies the onset of polarization as well as clearly indicates the existence of PNRs \cite{chen2013effectively,chen2015negative,cross1987relaxor}. 
\begin{figure}
\includegraphics[scale=0.4]{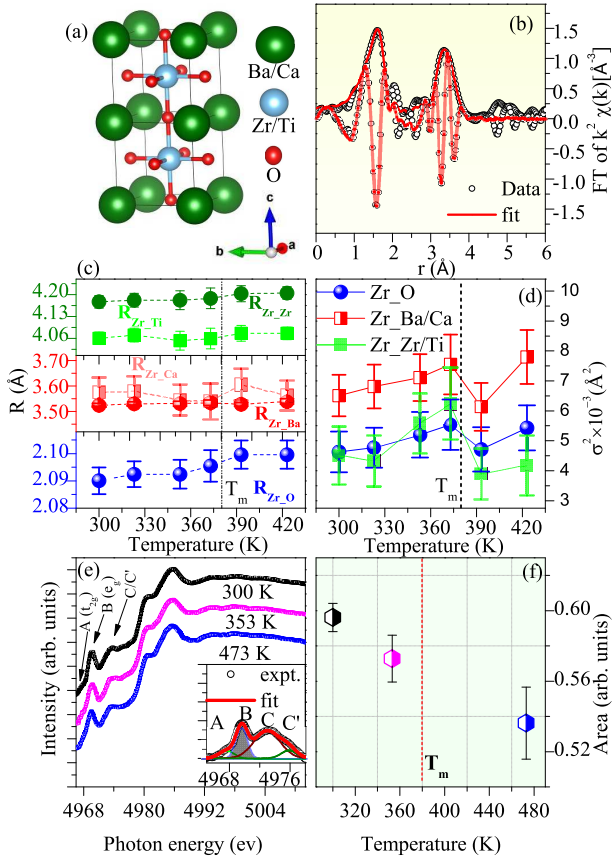}
\caption{\label{fig:epsart} (a) Standard perovskite model, which is used in the EXAFS fit. (b) Data and fit of the modulus and imaginary part of the Fourier transform in \textit{r} space at room temperature. (c) Temperature dependent coordination distances and (d) mean-square relative displacement (msrd) parameters involved in the fitting of the Zr K-edge EXAFS. (e) The Ti K-edge XANES spectra at different temperatures; the inset shows fitting of the preedge peak at room temperature. (f) Temperature dependence of the integrated intensity of feature B (\textit{e$_g$}) in (e).}
\label{EXAFS_XANES_New}
\end{figure}
Motivated by the above discussed results about evidences of PNRs, now we try to figure out the interplay between the Zr local structure and the PNRs.  The Zr local structure are investigated by performing the Zr K-edge EXAFS measurements at selected temperatures across T$_{m}$ and the results are shown in the Fig.~\ref{EXAFS_XANES_New}. We have used standard perovskite model (see Fig. ~\ref{EXAFS_XANES_New}(a)) with linear Zr-O-Zr, Zr-O-Ti links during fitting as has been used earlier in BZT compositions \cite{levin2011local}. During temperature dependent EXAFS analysis we have utilized the stoichiometric ratio of Ba/Ca and Ti/Zr obtained from fit at 300 K, which are found nearby nominal composition values (see details of the EXAFS analysis in \cite{supp1}). The fitted Zr K-edge EXAFS spectra at room temperature is shown in Fig. ~\ref{EXAFS_XANES_New}(b). From the results of first coordination shell we have found that Zr-O bond distance increases significantly with temperature (see Fig. ~\ref{EXAFS_XANES_New}(c)). The Zr-Ba \& Zr-Ca interatomic distances are not same at room temperature. As temperature increases both the parameters (R$_{Zr-Ba}$, R$_{Zr-Ca}$) come closer, while approaching T$_m$. The Zr-Ba interatomic distance ($\sim$3.54 $\AA$) is found significantly lesser than one in BaZrO$_3$ ($\sim$3.64 $\AA$) \cite{laulhe2006exafs} i.e., the ZrO$_6$ octahedra are not in an ideal cubic environment. Additionally, the Ti-O bond distance ($\sim$ 1.97 $\AA$) is found to be shorter compared to the Zr-O bond distance ($\sim$ 2.09 $\AA$). 
In order to better understand the local picture, we are turning our attentions to the mean square relative displacement (msrd) parameters as it contains information about thermal disorder as well as static disorder of all scattering paths involved. Fig. ~\ref{EXAFS_XANES_New} (d) shows evolution of msrd ($\sigma^{2}_{Zr-O}$) with temperature. The $\sigma^{2}_{Zr-O}$ at room temperature is higher in this system as compare to that of BaZrO$_3$ \cite{laulhe2006exafs}. So, here, the Zr-O bonds contain an additional disorder other than thermal disorder (Einstein's model) \cite{fornasini2015exafs, beni1976temperature}. Earlier \cite{laulhe2006exafs}, in case of BaZr$_x$Ti$_{1-x}$O$_3$, presence of a temperature independent static disorder together with thermal disorder was attributed to the distortion of ZrO$_6$ octahedra or tiny  displacement of the Zr atoms. Additionally, at $\sim$ T$_m$, $\sigma^{2}_{Zr-O}$ shows anomaly which is not seen earlier in case of BZT. Again, below T$_C$ behavior of the $\sigma^{2}_{Zr-Ti/Zr}$ with temperature is more steeper compared to the $\sigma^{2}_{Zr-O}$ $\&$ $\sigma^{2}_{Zr-Ba/Ca}$. This indicates modifications in linkage between the Zr-O and the O-Ti bonds due to decreasing covalent character of the Ti-O bonds. This is supported by a direct evidence of decreasing hybridization between the Ti and the O with temperature seen in the Ti K-edge X-ray absorption near edge structure (XANES) spectra (see Fig. ~\ref{EXAFS_XANES_New} (e,f)). The features A and B, in Fig. ~\ref{EXAFS_XANES_New} (e), are ascribed to t$_{2g}$ and e$_{g}$, respectively and C/C$\prime$ correspond to local Zr/Ti ratio surrounding absorber atom. The intensity of feature B (e$_g$), which is related to Ti-O hybridization, decreases with temperature (Fig. ~\ref{EXAFS_XANES_New}(f)), and confirms that with increasing temperature in this range strength of the Ti-O hybridization is weakening. Above T$_m$, $\sigma^{2}_{Zr-Ti/Zr}$ is more stable compared to other msrd parameters. Such temperature evolution of $\sigma^{2}_{Zr-Ti/Zr}$ suggests disorder along the Zr-O-Ti links which detains phonon activation. At around T$_m$, $\sigma^{2}_{Zr-O}$ also shows anomaly, indicating direct involvement of the Zr local structure on polarization. Based on our results from local structural analysis and the NPDF results by Pramanick\textit{et al.,} \cite{pramanick2018stabilization} we can say that presence of the Zr affects surrounding Ti ions with the tensile stress on it and can change the polarization (by slowing down dynamics of Ti$^{4+}$ ions).
\begin{figure}
\includegraphics[scale=0.295]{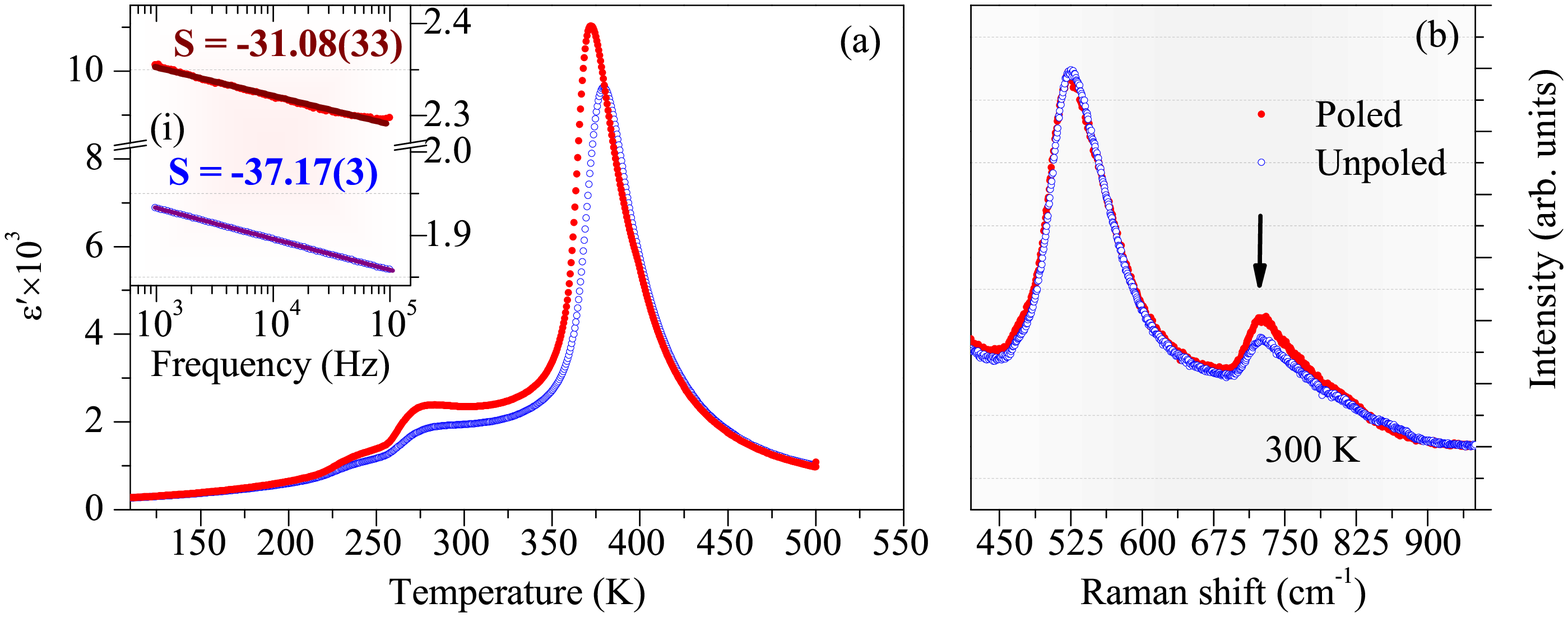}
\caption{\label{fig:epsart} (a) Temperature-dependent dielectric permittivity of poled and unpoled samples (at 1 kHz) taken during heating cycles. The inset shows the linearly fitted dielectric dispersion of poled and unpoled samples at room temperature. (b) Raman spectra (vertically translated for overlapping) of poled and unpoled samples at room temperature.}
\label{LT_PU_XRD_F}
\end{figure}  
\begin{figure*}
\includegraphics[scale=0.28]{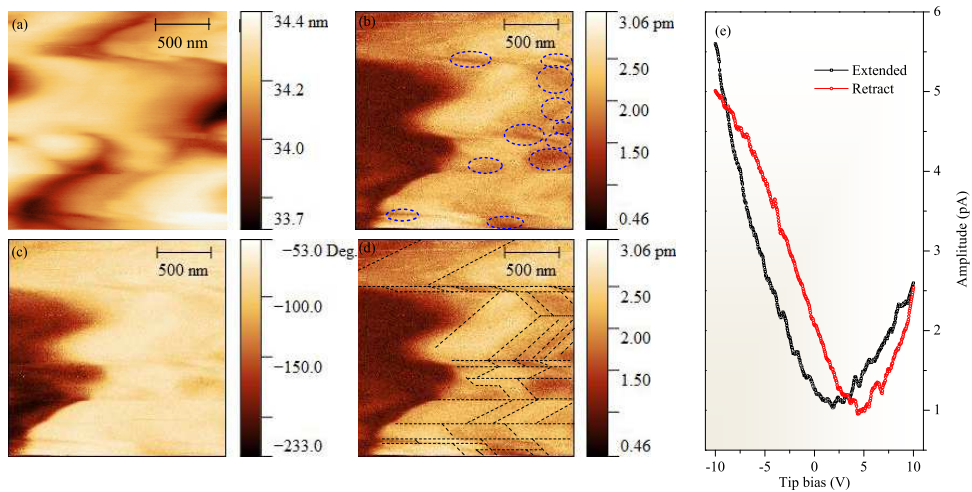}
\caption{\label{fig:epsart} (a) Topography, (b) piezoelectric amplitude, and (c) PRFM phase images of the same 2$\times$2 $\mu$m$^{2}$ scan area, at room temperature. The circles in (b) indicate different nanoscale domains. (d) Heirarchy of domains and subdomains marked by dashed lines and (e) local PRFM amplitude hysteresis loop with electric field.}
\label{PFM}
\end{figure*}
\begin{figure}
\includegraphics[scale=0.19]{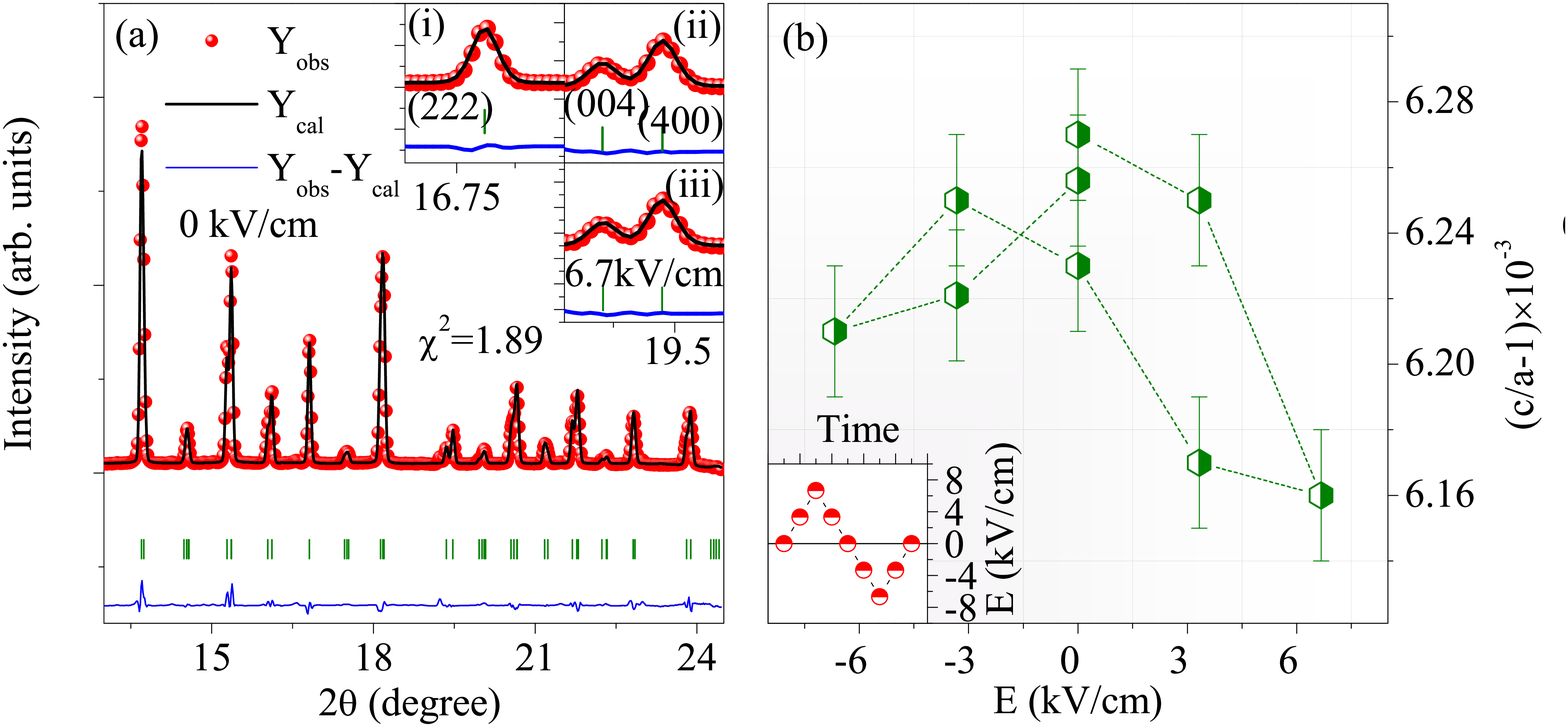}
\caption{\label{fig:epsart} (a) Rietveld refined X-ray diffraction pattern at zero electric field at room temperature. Insets (i) and (ii) in (a) show zoomed singlet (222)$_{pc}$ and doublet (400)$_{pc}$ peaks at zero field, and inset (iii) shows doublet (400)$_{pc}$ peak at  6.67 kV/cm. (b) Tetragonal distortion vs. electric field shows an inverted butterfly like feature. A bipolar electric field was applied as shown in the inset of (b).} 
\label{SXRD_F}
\end{figure}
In order to understand the effect of these local structural heterogeneities on long range polar order below T$_{m}$ we have also performed the dielectric measurements on poled sample and compared the results with that of unpoled sample. The dielectric dispersion or difference in permittivity at two frequencies is a measure of the degree of polar heterogeneity in a system \cite{groszewicz2016reconciling, adhikary2019random}. In general, the dielectric dispersion is less for long range ferroelectric materials and it increases for material that shows relaxor ferroelectric behavior which are complex perovskite (ABO$_3$) system with different cations present in the same site (A or B site). Relaxor ferroelectrics (i.e. BNT, BNT-BKT etc.) or canonical relaxors (i.e. PMN, PZN) can be transformed into long range polar order by the application of sufficiently high electric field, therefore dielectric dispersion as well as polar heterogeneity get decreased. As we see from EXAFS results that the studied composition exhibits significant disorder due to different sizes’ cations are present at both the A and the B sites, so a comparison of dielectric dispersion in poled and unpoled sample can directly tell about the degrees of polar heterogeneities. 
The inset of Fig. ~\ref{LT_PU_XRD_F}(a) shows that the dielectric dispersion decreases significantly for poled sample therefore it is inferred that the degree of polar heterogeneity is reduced after poling. These observation of presence of polar heterogeneities in the unpoled sample can be correlated to structural heterogeneities observed in the EXAFS experiment.

Surprisingly, dielectric permittivity enhances in case of poled sample which can be visibly seen from Fig. ~\ref{LT_PU_XRD_F}(a). Generally, by applying electric field on ferroelectric domains, polarization coherence increases along field direction and the sample becomes less susceptible to smaller ac field and thereby dielectric permittivity reduces \cite{yadav2017coexistence,koa2004dielectric}. Similar is true in case of canonical relaxors (PMN, PZN) \cite{zhang2000influences} or in the nonergodic phase of relaxor ferroelectrics (e.g. BNT based systems) below T$_{m}$ \citep{adhikary2019random,rao2013local}. However, field induced real permittivity enhancement is reported to exist below T$_{m}$ in case of PMN-PT, and that has been attributed to synergistic interaction among ferroelectric domains and PNRs \cite{li2016origin}, as may be the case here too. We also performed Raman measurements on the poled and the unpoled samples (see Fig. ~\ref{LT_PU_XRD_F}(b)). The Raman spectra of these samples match well with the previous reports \cite{zhang2014phase,singh2017investigation,datta2020adaptive}. The Raman spectra of this sample as well as other samples of BZT-xBCT series (x = 0.3, 0.5, 0.6, 0.7) are compared and presented in supplementary \cite{supp1}. The Raman mode at $\sim$ 522 cm$^{-1}$ (see Fig. ~\ref{LT_PU_XRD_F}(b) and Fig. S4 in \cite{supp1})have dominant A$_1$(TO) behavior and is related to BO$_{6}$ octahedra \cite{buscaglia2014average}.  The Ramam mode A$_1$(LO)/E(LO) at $\sim$ 720 cm$^{-1}$ is key signature related to the long range ferroelectricity \cite{keswani2018role,pasha2007situ}. Both of these modes are present in all BaTiO$_3$-based systems. The intensity of the Raman mode (at $\sim$ 720 cm$^{-1}$) enhances after poling and suggests that long range ferroelectricity has indeed enhanced after poling \cite{buscaglia2014average,keswani2018role}. Therefore, both these measurements (dielectric and Raman) clearly suggest that there exist nanoscale polar inhomogeneities along with long range ferroelectricity below T$_{m}$ and reduces marginally in poled sample.  

In order to visualize these local structural heterogeneities we have also performed the domain imaging by PRFM. The PRFM is a technique based on the detection of local converse piezoelectric effect and this is promising for imaging fine scale domains \cite{urvsivc2019investigations,kim2015situ,otonicar2018multiscale}. Fig.~\ref{PFM} (a) shows the surface morphology of the polished sample by scanning a 2$\times$2 $\mu$m$^{2}$ region. Fig.~\ref{PFM} (b) $\&$ (c) show the piezoelectric amplitude and PRFM phase of the same scanned region. Ferroelectric domains of different orientations with different contrast can be clearly seen in Fig. \ref{PFM} (b) $\&$ (c) . Nanoscale domain configurations (marked by dot circles) are observed for this composition within the large ferroelectric domain (see Fig.~\ref{PFM} (b)). Fig.~\ref{PFM} (d) shows the hierarchical domain configuration and is similar to the earlier reports resulting due to the interplay of stress accommodation and polarization compensation \cite{yao2010hierarchical,wang2006hierarchical,fang2019multi,yao2011evolution}. Fig.~\ref{PFM} (e) shows the piezoelectric amplitude hysteresis loop which was obtained by applying electric field along the tip axis on selected region. The butterfly like piezoelectric amplitude hysteresis loop confirms the existence of well defined polarization along field direction. Historically, high piezoresponse is always associated with hierarchical domain structure or complex microstructure with a combination of microdomains and nanodomains as these domains can easily respond to the external stimulus\cite{hu2020ultra,wang2006hierarchical,khachaturyan2010ferroelectric}. The complex domain configuration of this sample is compatible with its high piezoresponse ($\sim$223 pC/N) \cite{hu2020ultra,khachaturyan2010ferroelectric}. 
   
Finally, for the sake of completeness we carried out X-ray diffraction (at $\sim$ 300 K, below T$_{m}$) in presence of field up to $\pm$ 6.7 kV/cm ($\sim$ 1.67E$_c$) and quantified the structural changes under electric field . Electric field was applied on sample at an angle of 45$^{\circ}$ with respect to the incident X-ray beam. This filters out effects of ferroelastic domain induced texturing and also assures reliability of the structural information through the Rietveld refinement \cite{hinterstein2011structural}. Fig. ~\ref{SXRD_F}(a) shows the fitted XRD pattern, which is in good agreement with the experimental data. Fig.~\ref{SXRD_F}(b) shows tetragonal distortion with electric field, though small. The observed tetragonal distortion reduction clearly indicates the electric field induced 90$^{\circ}$ domain wall motion by reduction of anisotropy. In the Pb-based compositions, tetragonal distortion reduces by enhancing the monoclinic phase fraction \cite{guo2000origin,hinterstein2011structural,liu2017critical}.  However, in the present case no traces of monoclinicity is observed. Zoomed portions around (004) and (400) under highest electric field is depicted in inset (iii) of Fig.~\ref{LT_PU_XRD_F}(a) and no extra features are observed, as is the case in Pb-based compounds. Our electric field dependent XRD results and conclusions  are well in agreement with those already reported on the same composition \cite{ehmke2014resolving,tutuncu2014domain}   
 
\section{Discussion}
BZT-60BCT is found to exhibit a weaker dielectric dispersion at around T$_{m}$ compare to the canonical relaxors PMN and PZN; however, other characteristics of relaxor type exist, e.g., a temperature of the real permittivity maxima ($\sim$380 K) significantly higher than that of the imaginary permittivity maxima ($\sim$375 K), deviation from the C-W law at a temperature (at $\sim$ 477 K) much higher than T$_m$, and the presence of a slim hysteresis loop above T$_m$. The diffusion exponent of this system ($\gamma$=1.52) suggests that this system exhibits diffused ferroelectric phase transition, an intermediate between classical ferroelectric ($\gamma$=1) and canonical relaxor ($\gamma$=2) \cite{nuzhnyy2012broadband}. Therefore, this system is analogically similar to PMN-PT and PZN-PT compositions near the MPB \cite{noblanc1996structural,li2016origin}. The dielectric anomalies (deviation from C-W law and structural phase transition around T$_m$) are successfully justified by temperature dependent XRD. The appearance of SVFS at $\sim$477 K which is associated with the onset of the PNRs is in accordance with other perovskite (ABO$_{3}$) ferroelectrics \cite{chen2015negative} in which polarization developments also take place due to the relative displacements of the cations and anions.

The origin of nanodomains here may relate due to the inhomogeneities in the local structure and in strain (cations with different sizes) which break the long-range polar order. The compositions studied here are perfect candidate for such a scenario because these are highly disordered compound as it involves Zr$^{4+}$ (ionic radii \textit{r} $\sim$ 0.72$\AA$) substitution in place of Ti$^{4+}$ (\textit{r} $\sim$ 0.61$\AA$) and Ca$^{2+}$ (\textit{r} $\sim$ 1.34$\AA$) in place of Ba$^{2+}$ (\textit{r} $\sim$ 1.61$\AA$) and thus large ionic radii \cite{shannon1976revised} differences at both sites. 
Weakening of long range polar-order as a function of the Zr and Ca content in these compositions is experimentally evidenced by our Raman measurements, for which we have compared the four different compositions of BZT-xBCT (x = 0.3, 0.5, 0.6, 0.7) \cite{supp1}. The A$_1$(TO) phonon mode \cite{keswani2018role,buscaglia2014average}, related to polar Ti-O vibrations (at $\sim$ 270 cm$^{-1}$), clearly becomes weaker and broader for Ca$^{2+}$ rich BZT-BCT systems.
 
Unlike Pb-based relaxors in which Pb positional disorder plays an important role, in BZT-BCT it has been reported \cite{datta2020adaptive} that at the A site (Ba/Ca) there is no significant disorder. However, from our Zr local structural studies we have found a significant difference between R$_{Zr-Ba}$ and R$_{Zr-Ca}$ which together with our Raman results highlights that the A site (Ba/Ca) also contributes to local structural heterogeneities. Recent first-principle studies \cite{amoroso2018first,amoroso2019interplay} also highlighted the steric effect induced B-site to A-site ferroelectricity transformation by incorporating lower ionic radii Ca in Ba sites. However, to shed light on the peculiar role of Ca more experimental investigations are necessary. 

In PZT ceramics relaxor behavior is absent, although the same Ti and Zr cations are present in the B sites as in BZT. This is mainly because, the A site cations possess a contrasting character.  The stereochemically active lone pairs of Pb in covalently bonded Pb-O result in large off centering in their respective oxygen dodecahedra, whereas ionic-type bonding between the Ba-O results in almost no shifting of Ba. In Pb-based system, Pb positional disorder couples differently with different BO$_6$ octahedra; strong correlations between the cations make PbTiO$_3$ ferroelectric and PbZrO$_3$ antiferroelectric \cite{kuroiwa2001evidence,tagantsev2013origin,cao2004local}. In BZT-BCT substitution of Ca at the Ba sites, however, increases the A site bond strain due to a large ionic radii mismatch between Ba and Ca; however, the effect of the Ca on the BO$_6$ octahedra is not as prominent as that of Pb. In BaTiO$_3$-based systems (BT, BZT, BST) polarization appears mainly due to the Ti$^{4+}$ and the effect of dopant ions in BaTiO$_3$ system exhibits diffuse scattering by setting up random strain fields generating frustrated PNRs \cite{liu2007structurally}, whereas PZT does not show diffuse scattering \cite{phelan2014role}. Moreover, in BaTiO$_3$-based relaxors B site substitution has a more important role in setting up local strains by decreasing transverse correlations and creating frustration during long-range polar order \cite{liu2007structurally}. Fanoresonance and the thermally activated THz relaxation mode \cite{wang2014fano} due to the difference in off-centering of Zr$^{4+}$ and Ti$^{4+}$ is attributed as the cause of PNRs or relaxor behavior in BZT \cite{akbarzadeh2012finite}. The Ti$^{4+}$ dynamics are reported to exist even in the non-ergodic relaxor phase of BZT \cite{petzelt2014broadband,petzelt2015lattice}.
    
Generally for non-ergodic relaxors (below T$_m$) dielectric dispersion is larger (compare to long range ferroelectrics), and application of a sufficiently high electric field results in a reduction of dielectric dispersion. Dispersion in the permittivity data signifies the degree of polar inhomogeneity \citep{adhikary2019random, groszewicz2016reconciling}. Here, in this system, dispersion in the real permittivity is also present well below T$_m$. The slopes calculated from dielectric permittivity variation with frequency (inset in Fig.~\ref{LT_PU_XRD_F}(a)) suggest that significant polar inhomogeneity is present in case of the unpoled specimen and it decreases only marginally after electric poling \cite{supp1}. Normally, the presence of PNRs below T$_{m}$ is expected to show nonergodicity \cite{bokov2012dielectric} and application of electric field results in the reduction of real permittivity, which here is counter-intuitive\cite{adhikary2019random}. Further, both, PNRs and ferroelectric domains individually are reported to show enhancement of polarization coherence after poling thereby a reduction in the dielectric constant \cite{koa2004dielectric, li2017contributions, li2018local}. In contrary, Li \textit{et al.,} \cite{li2016origin} highlighted dielectric constant enhancement after poling as the cornerstone of ultrahigh piezoelectricity. The temperature evolution of structural transformations of PNRs at lower temperature, and colinear arrangement of PNRs at higher temperature were attributed  to dielectric as well as piezo amplification in PNR-ferroelectric composites \cite{li2016origin}. In our case also, in the poled sample real permittivity increases significantly ($\sim$23$\%$). We speculate that similar kinds of interactions are involved in this system, as explained by Li \textit{et al} \cite{li2016origin}. Our PRFM measurements show the presence of polar nanodomains at room temperature and support our conjecture.   
\section{Conclusions}
In summary, by combining a set of local and average structural studies on a tetragonal BZT-BCT ceramic we have presented the relationship between local structural heterogeneities in it and its macroscopic dielectric, ferroelectric and piezoelectric properties. Temperature dependent XRD confirms tetragonal to cubic phase transition at $\sim$ T$_{m}$ (380 K). Furthermore, the onset of spontaneous volume ferroelectrostriction and deviation from the C-W law (at T$_{B}\sim$477 K), slim hysteresis and dielectric dispersion confirm the presence of PNRs from 477 K in this composition. The Zr local structural investigations confirm that the ZrO$_6$ \& TiO$_6$ octahedra retain the individuality of their respective parent compounds (i.e, BaZrO$_3$ and BaTiO$_3$) and ZrO$_6$ octahedra are found to have direct involvement during polarization developement. The larger ZrO$_6$ octahedra  create tensile stress on surrounded TiO$_6$ octahedra and the presence of static disorder in Zr-O bonds produces local structural heterogeneities; during polarization development the ZrO$_6$ octahedra distort. Local structural heterogeneities try to break the long-range polar order by forming nanoscale hierarchical domains. These domains are evidenced by PRFM and are supposed to be responsible for the enhancement of real permittivity ($\sim$23\%) after poling. Additionally, high resolution SXRD confirms also the electric field induced reduction of tetragonal distortion in this ceramic.  
\section*{Acknowledgments}
K.D., A.A. and D.K.S gratefully acknowledge the financial support from the Department of Science and Technology (DST) in India through the India-DESY collaboration for performing experiments at PETRA-III, DESY. The authors acknowledge S. Yadav for help with P-E loop measurements, and A. Yadav for help with EXAFS data analysis. D. Reuther is acknowledged for setting up the voltage source used in the electric field dependent XRD measurements. C.R. acknowledges the support from the project CALIPSOplus under  Grant Agreement No. 730872 from the EU Framework Programme for Research and Innovation HORIZON 2020.  
\bibliography{KD_60}
\end{document}